\documentclass[a4paper
]{article}
\usepackage{graphicx}
\usepackage[]{authblk}
\usepackage[T1]{fontenc}
\usepackage[utf8]{inputenc}
\pdfoutput=1
\usepackage{epstopdf}
\usepackage{amsmath}
\usepackage{subfig}
\usepackage{float}
\usepackage{indentfirst}
\usepackage{dcolumn}
\usepackage{bm}
\usepackage[letterpaper,includehead,top=1cm,left=3cm,right=3cm,headsep=1.5cm,headheight=0.0cm,bottom=2.5cm, footskip=1.0cm]{geometry}

\title{Instability in Shocked Granular Gases}
\author[1]{Nick Sirmas\thanks{nsirmas@uottawa.ca}}
\author[2]{Sam Falle}
\author[1]{Matei Radulescu}
\affil[1]{Department of Mechanical Engineering, University of Ottawa, ON, K1N6N5, Canada}
\affil[2]{Department of Applied Mathematics, University of Leeds, Leeds, LS2 9JT, UK}
\begin{document}
\maketitle


\begin{abstract}
Shocks in granular media, such as vertically oscillated beds, have been shown to develop instabilities. Similar jet formation has been observed in explosively dispersed granular media. Our previous work addressed this instability by performing discrete-particle simulations of inelastic media undergoing shock compression. By allowing finite dissipation within the shock wave, instability manifests itself as distinctive high density non-uniformities and convective rolls within the shock structure. In the present study we have extended this work to investigate this instability at the continuum level. We modeled the Euler equations for granular gases with a modified cooling rate to include an impact velocity threshold necessary for inelastic collisions. Our results showed a fair agreement between the continuum and discrete-particle models. Discrepancies, such as higher frequency instabilities in our continuum results may be attributed to the absence of higher order effects.
\end{abstract}

\section{Introduction}

Experiments have shown that shocks in granular media can become unstable. For example, granular media in vertically oscillating beds have been shown to become unstable \cite{Bizon1998}. Another example is in similar jet formation observed in explosively dispersed granular media \cite{Frost2012}. 

Previous studies have addressed the unique structure of shock waves through granular media, although instabilities had not been identified \cite{Goldshteinetalch31996, Kamenetsky_etal2000}. These structures were identified by studying the problem of piston propagated shock waves. In this structure, a piston causes a shock wave to traverse through a granular media, which causes the granular temperature to increase. Due to the inelasticity and increased rate of the collisions within this ``fluidized'' region, the granular temperature decreases and the density increases. Eventually, the density is high enough that the collisions have subsided, characterized as the ``frozen'' equilibrium region.

The structure of granular shock waves is similar to that of strong shock waves through molecular gases that undergo strong relaxation effects \cite{Zeldovich&Raizer1966}; whereby the difference exists in the equilibrium region, where a fraction of translational kinetic energy is conserved. Experiments have shown that such shock waves can become unstable due to endothermic reactions (eg. ionization \cite{Grunetal1991, Griffithetal1976} and dissociation \cite{Griffithetal1976}). Although there is significant experimental evidence showing instability in relaxing shock waves, the mechanism controlling the instability is not well understood. 

Shock waves through granular media and molecular gases undergoing endothermic reactions share a common feature, where the shock is characterized by dissipative collisions \textit{within} the shock structure. Therefore, investigating the controlling mechanism of instability in granular gases may have direct bearing on the understanding of instabilities seen in molecular gases.

Previously, we have shown shock instability in a system of inelastic disks (2D), with collisions of disks modelled deterministically \cite{RadulescuSirmas2011}. Introducing an activation energy for which disks collide inelastically yields the formation of high density non-uniformities and convective rolls within the relaxing region of the shock structure. By studying the time evolution of the material undergoing shock compression and further relaxation, we find that the granular gas develops the instability on the same time scales as the clustering instability in homogeneous granular gases. Thus confirming that the clustering instability is the dominant mechanism controlling instability in our model \cite{SirmasTBD}.

Previous studies investigating Faraday instabilities in granular media have shown a similar jetting instability \cite{Carrilo2008}. These studies have been done with the same molecular dynamics model \textit{and} with continuum modelling of the Navier-Stokes granular equations. Both of these models are in good agreement, reproducing the Faraday instability. 

Thus, we pose the following question pertaining to our investigation: \textit{Can we observe a similar shock instability in granular gases at the continuum level?} Specifically, we wish to examine whether the instability can be seen using a simple hydrodynamic description of granular media, absent of higher order viscous effects.

For a granular system with a constant coefficient of restitution, the hydrodynamic descriptions are well understood \cite{Brilliantov&Poschel2004}. However, in our study we introduce an activation energy necessary for inelastic collisions. This requires modifications to the general transport coefficients, using methods from traditional kinetic theory of monatomic gases. 

This paper is organized as follows. First, we describe the model that is used to study the stability of shock waves through granular gases, with the presence of an activation energy. Secondly, we describe the numerical methods we use for the molecular and continuum models. Specifics are outlined for the modifications made to the governing equations. Finally, the shock structures from both methods are compared.

\section{Problem Overview}

The medium we investigate is a system of colliding hard disks. Each binary collision is elastic, unless an \textit{activation} threshold is met. Quantitatively, collisions are inelastic if the impact velocity (normal relative velocity) exceeds a velocity threshold $u^*$, a classical activation formalism in chemical kinetics. If the collision is inelastic, the disks collide with a constant coefficient of restitution $\varepsilon$.

The system we study is a classical shock propagation problem, whereby the motion of a suddenly accelerated piston driven into a thermalized medium drives a strong shock wave. The driving piston is initially at rest and suddenly acquires a constant velocity $u_p$.  This model allows for the dissipation of the non-equilibrium energy accumulated within the shock structure, which terminates once the collision amplitudes fall back below the activation threshold.  In this manner, the activation threshold also acts as a tunable parameter to control the equilibrium temperature in the post shock media.

\section{Numerical Models}
\subsection{Molecular Dynamics Model}
The MD simulations reconstruct the dynamics of smooth inelastic disks, calculated using the Event Driven Molecular Dynamics technique \cite{Alder&Wainright1959}.  We use the implementation of P\"{o}schel and Schwager \cite{Poschel&Schwager2005}, that we extended to treat a moving wall. The particles were initialized with equal speed and random directions.  The system was let to thermalize and attain Maxwell-Boltzmann statistics.  Once thermalized, the piston started moving with constant speed. 

The initial packing fraction of the disks was chosen to be $\eta_1=(N\pi d^2)/4A=0.012$, where $N\pi d^2/4$ is the specific volume (area) occupied by $N$ hard disk with diameter $d$, and domain area $A=L_x\times L_y$ ; the initial gas is thus in the ideal gas regime \cite{Sirmasetal2012}. 

All distances and velocities are normalized by the initial root mean squared velocity $u_{rms_o}$ and the diameter $d$ of the disks, respectively. Thus fixing the time scaling $\overline{t}=t/(\frac{d}{u_{rms_o}})$. The initial domain size was $A=4550d\times455d$,  occupied by 30,000 hard disks.

\subsection{Continuum Model}
In the continuum model, we consider a two-dimensional granular gas, by modelling a system of smooth inelastic disks. For such a system, the Euler hydrodynamic equations for mass, momentum and energy take the form:
\begin{eqnarray}
\frac{\partial \rho}{\partial t}+\vec{\nabla}\cdot\left(\rho\vec{u}\right)&=&0\notag\\
\frac{\partial \rho\vec{u}}{\partial t}+\vec{\nabla}\cdot\left(\rho\vec{u}\vec{u}\right)&=&-{\nabla}p\\
\frac{\partial E}{\partial t}+\vec{\nabla}\cdot\left(\vec{u}(E+p)\right)&=&\zeta\notag
\end{eqnarray}
where $E=\rho T+\frac{1}{2}\rho \vec{u}^2$ is the total energy density, in terms of density $\rho$, granular temperature $T$ and velocity $\vec{u}$.
For granular systems, the hydrostatic pressure $p$ can be approximated as:
\begin{equation}\label{EOS1}
p=\rho e\left[ 1+(1+\varepsilon)\eta g_2(\eta)\right]
\end{equation}
where $e=T$ is the internal energy in 2D and $g_2(\eta)=(1-(7/16)\eta)((1-\eta)^2)$ is the pair correlation function for a system of hard disks \cite{Torquato1995}. 

The cooling coefficient $\zeta$ for constant $\varepsilon$ may be written as:
\begin{equation}\label{zeta_nocorrect}
\zeta=-\frac{4}{d\sqrt{\pi}}\left(1-\varepsilon^2\right)\rho T^{3/2}\eta g_2(\eta)
\end{equation}
\subsubsection{Modification of Cooling Rate}

In our model, inelastic collisions occur for a fraction of collisions, making the cooling rate from \eqref{zeta_nocorrect} invalid. Therefore, a modification to the cooling rate is necessary to account solely on energy losses from activated inelastic collisions.

To adjust the cooling rate, one must examine the energy involved in collisions. 
To do so, we first look at the rate of binary collisions per unit volume \cite{Vincenti&Kruger1975}:
\begin{equation}\label{distribution}
n^2 d \frac{m}{2 kT}\exp\left\{\frac{m g^2}{4kT}\right\}g^2 \cos \psi dg d\psi
\end{equation}
This term gives the rate of binary collisions of a system of disks of mass $m$ with a number density $n=N/A$ that have a relative speed in the range of $g$ to $g+dg$, with an angle between the relative velocity and the line of action in the range of $\psi$ to $\psi+d\psi$. Along the line of action, the relative velocity is $u_n=g\cos\psi$. Multiplying \eqref{distribution} by $u_n^2=(g\cos\psi)^2$, and integrating over a range of $u_n$, one recovers the energy along the line of action for collisions with impact velocities within this range. Integrating with $u_n$ from 0 to $\infty$ recovers the energy directly applicable to \eqref{zeta_nocorrect}. Integrating $u_n$ from $u^*$ to $\infty$ yields the energy seen along the line of action for impact velocities exceeding $u^*$.  

To modify the cooling rate \eqref{zeta_nocorrect}(that considers all collisions inelastic), we calculate the ratio of the average energy involved in activated collisions to the energy of all collisions. This ratio is:

\begin{eqnarray}
\frac{\langle{u_{n}^2}\rangle(u_n>u^*)}{\langle{u_{n}^2}\rangle(u_n>0)}=\exp\left\{\frac{m {u^*}^2}{4kT}\right\}\left(1+\frac{m {u^*}^2}{4kT}\right)
\end{eqnarray}
Since $\frac{1}{2}{u^*}^2/u_{rms}^2=E_a/T$ for disks of equal mass \cite{Vincenti&Kruger1975}, multiplying this ratio with the cooling rate \eqref{zeta_nocorrect} yields the following modified cooling rate:
\begin{equation}
\zeta^*=-\frac{4}{d\sqrt{\pi}}\left(1-\varepsilon^2\right)\rho T^{3/2}\eta g_2(\eta)\exp \left\{ \frac{E_A}{T}\right\}\left(1+\frac{E_A}{T}\right)\\
\end{equation}

This cooling rate is validated by comparing the evolution of granular temperature obtained via MD for a homogeneous cooling granular gas with the inclusion of an activation threshold.

\subsubsection{Details of Hydrodynamic Solver}

The software package \textit{MG} is used to solve the governing equations of the described system. The \textit{MG} package utilizes a second order Godunov solver with adaptive mesh refinement. 

The hydrodynamics of the granular gas are solved by investigating the flow at the piston frame of reference. A reflective wall boundary is implemented on the left (piston face) with flow travelling towards the wall at constant velocity $u_p$. The upper and lower boundaries have reflective wall boundary conditions, and the right boundary has free boundary conditions. In order to compare with the MD results a domain height of $455d$ was used. A resolution of $\Delta x=d$ was found to be sufficient for the model.

In order to investigate the unstable shock structure, the initial and incoming flow density are perturbed. Random density perturbations are implemented with a variance of $10\%$, which are applied in perturbed cells with area $11.375d\times 11.375d$ (40 cells high in the current domain). The size of the perturbed cells is chosen such that the frequency of instability is not fixed, yet sufficient to avoid numerical diffusion of the flow. The variance is taken in accordance with deviation associated with the MD model with equal dimensions and packing factor.

\section{Results}

\begin{figure}[h]
\captionsetup[subfigure]{labelformat=empty, oneside,margin={-1cm,0cm}} 
\subfloat{\includegraphics[width=0.19\linewidth]{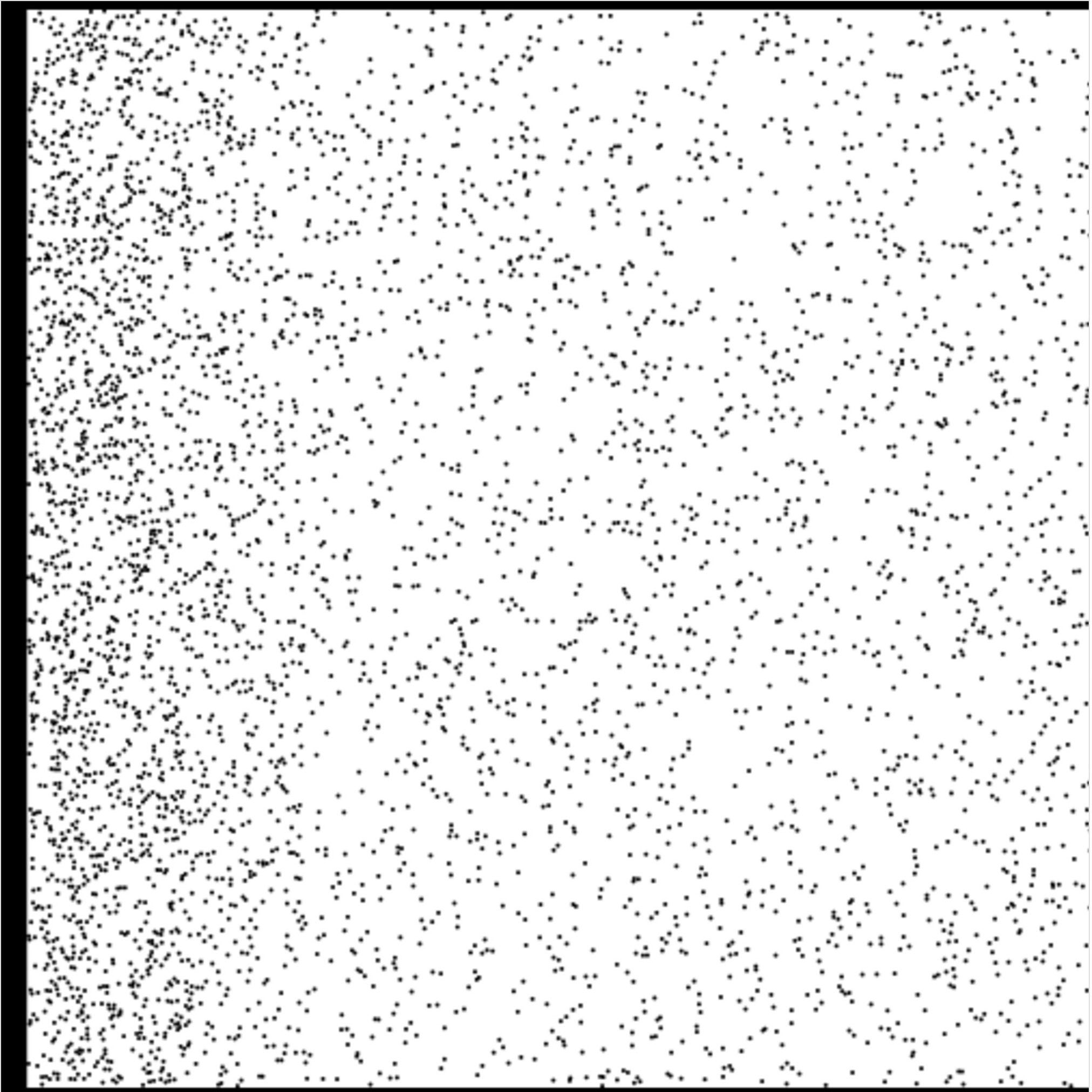}}
\subfloat{\includegraphics[width=0.19\linewidth]{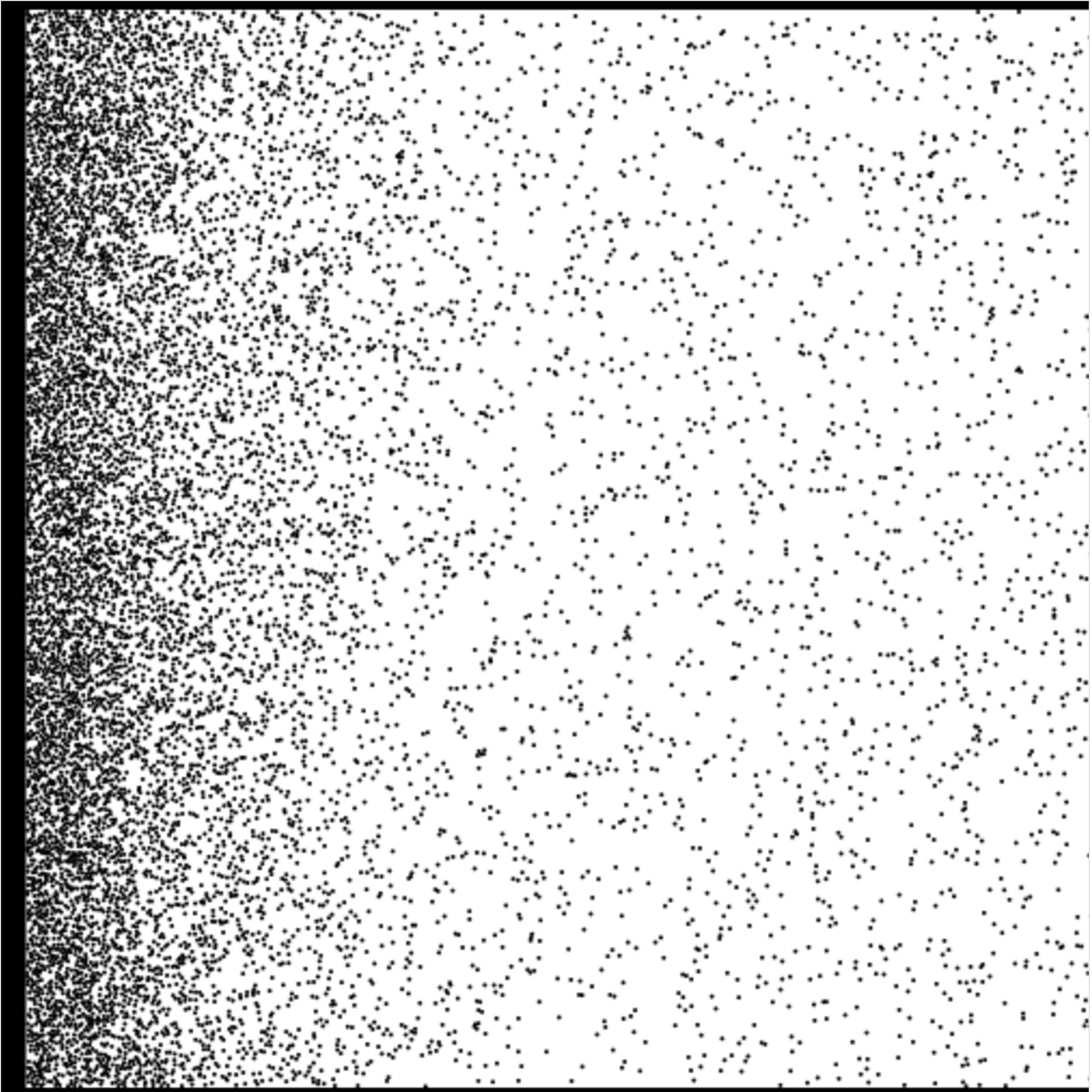}}
\subfloat{\includegraphics[width=0.19\linewidth]{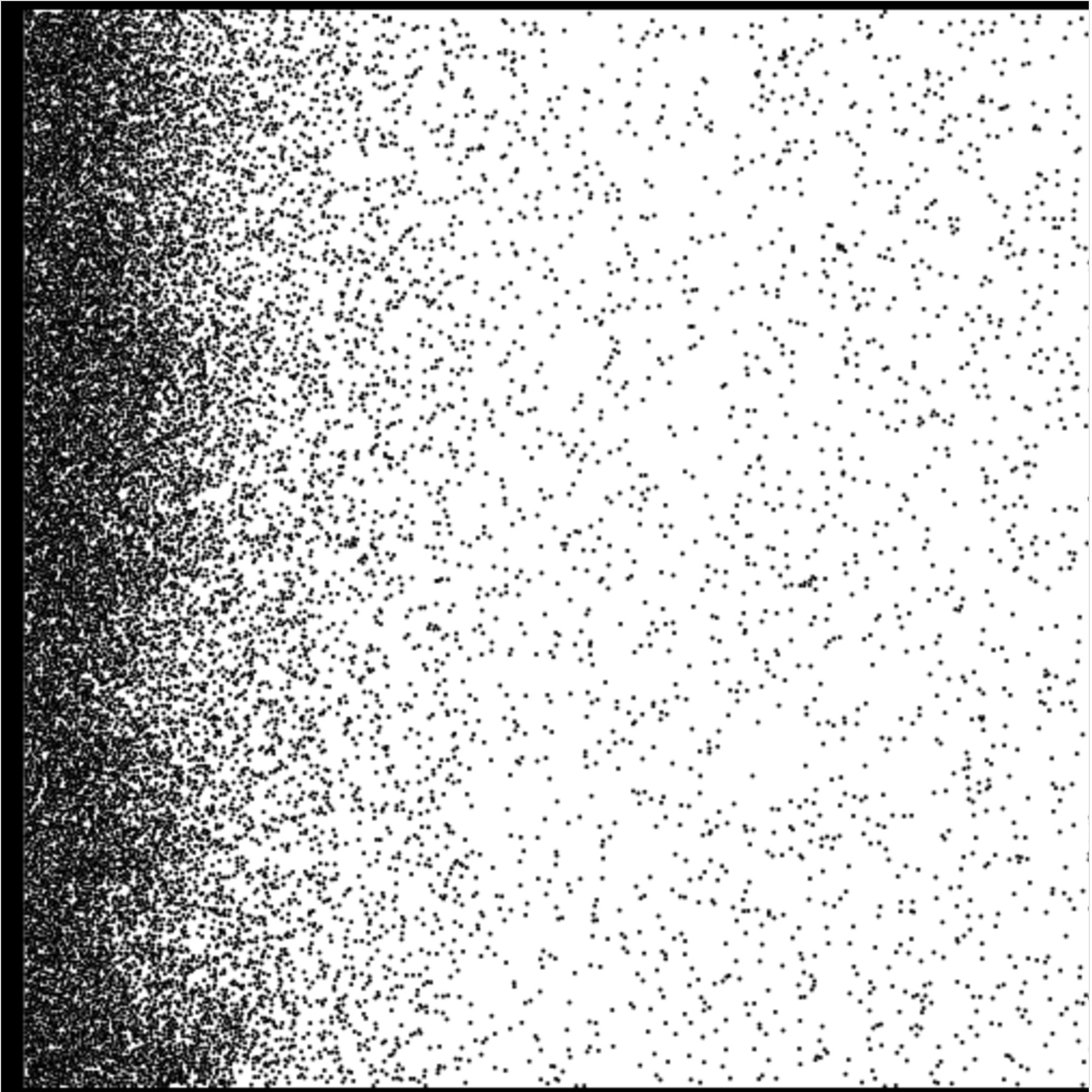}}
\subfloat{\includegraphics[width=0.19\linewidth]{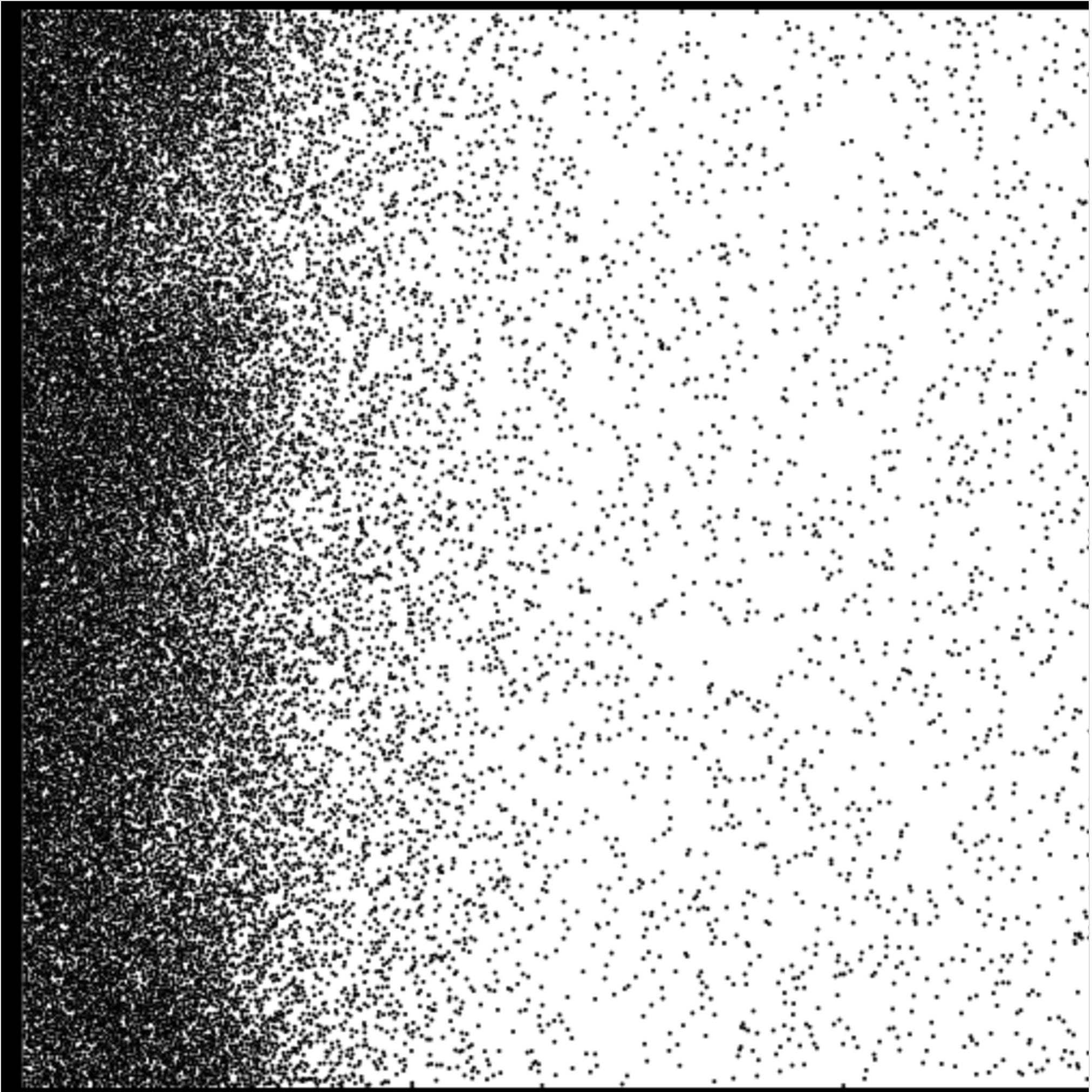}}
\subfloat{\includegraphics[width=0.19\linewidth]{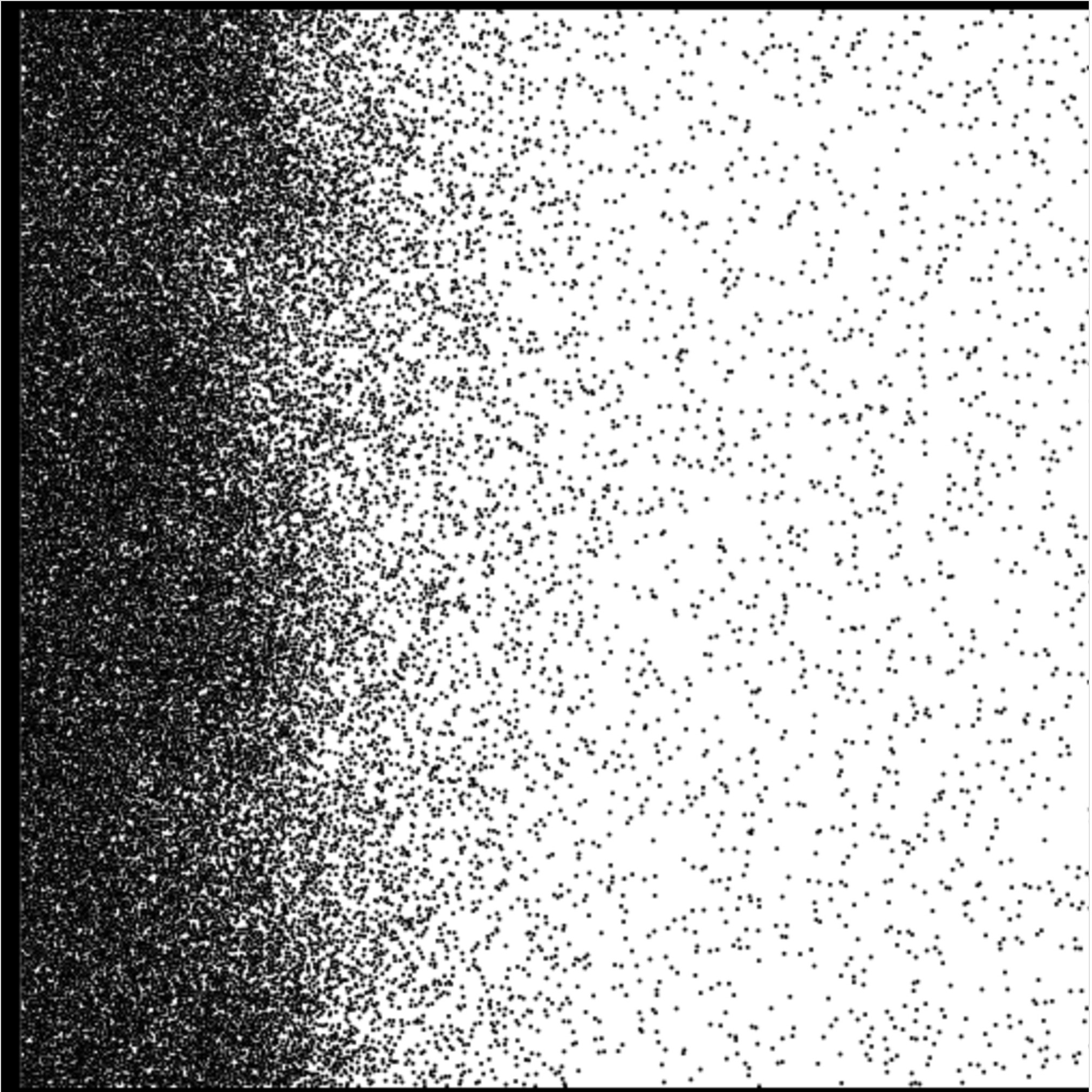}}\\
\vspace*{-5pt}\hspace*{-4pt}
\subfloat[$\overline{t}=45.5$]{\includegraphics[width=0.25\linewidth]{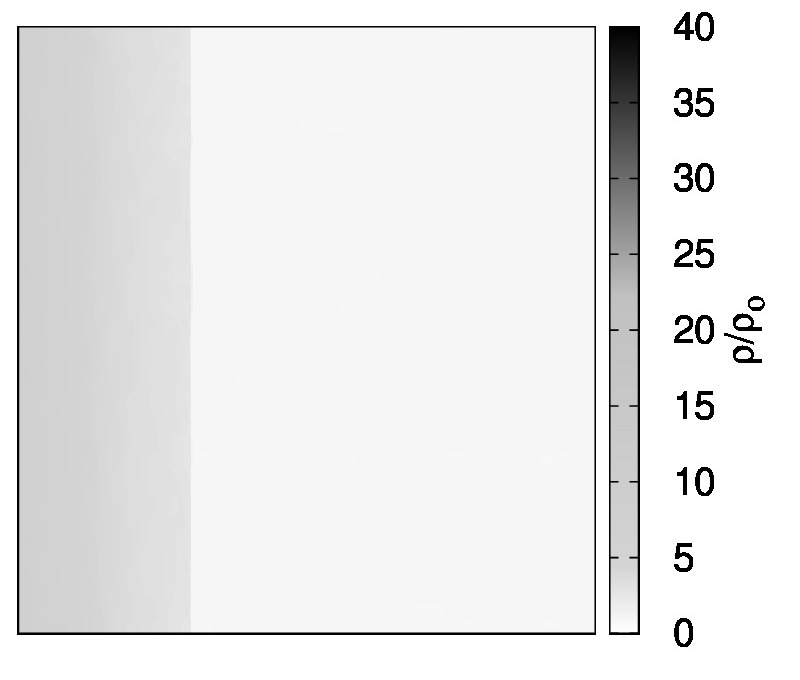}}\hspace*{-28.5pt}
\subfloat[$\overline{t}=91.0$]{\includegraphics[width=0.25\linewidth]{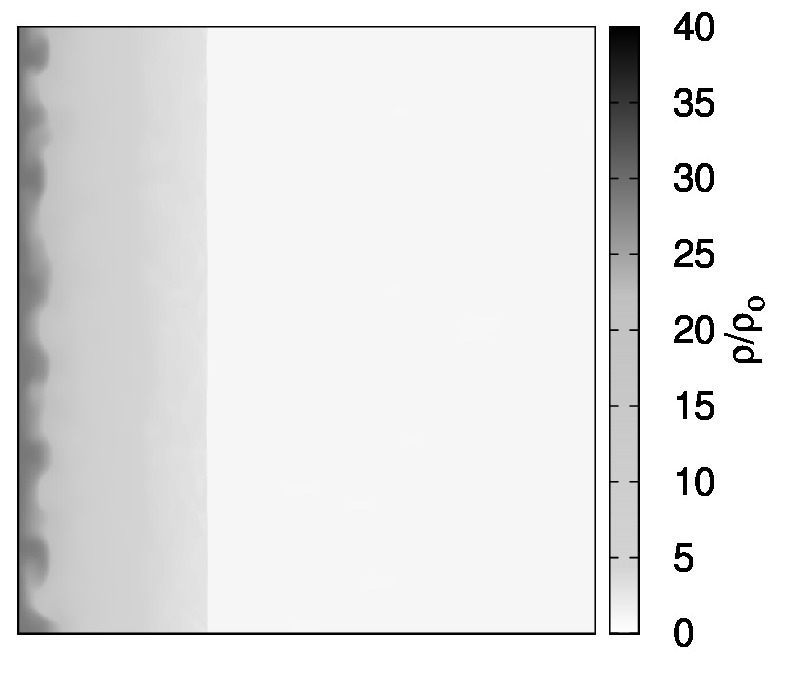}}\hspace*{-28.5pt}
\subfloat[$\overline{t}=136.5$]{\includegraphics[width=0.25\linewidth]{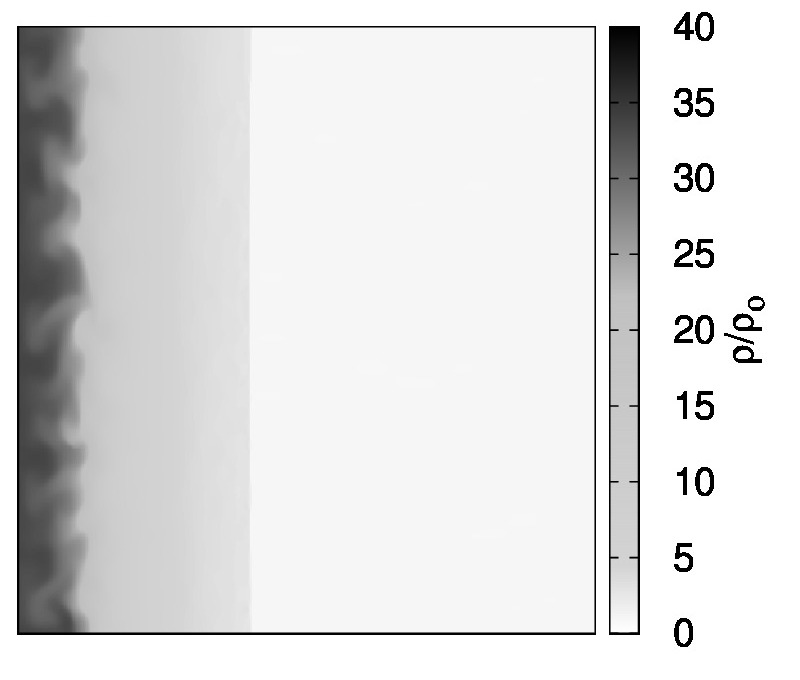}}\hspace*{-28.5pt}
\subfloat[$\overline{t}=182.0$]{\includegraphics[width=0.25\linewidth]{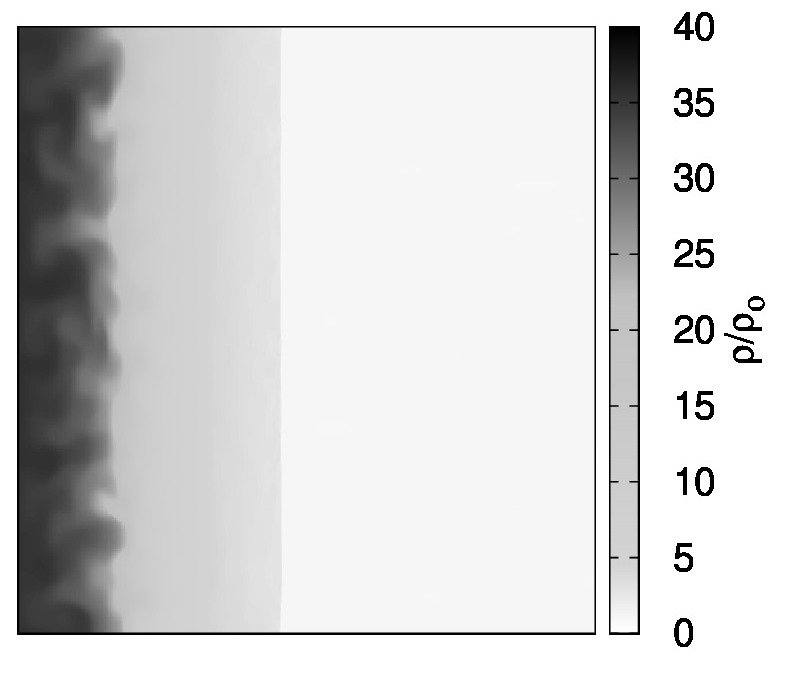}}\hspace*{-28.5pt}
\subfloat[$\overline{t}=227.5$]{\includegraphics[width=0.25\linewidth]{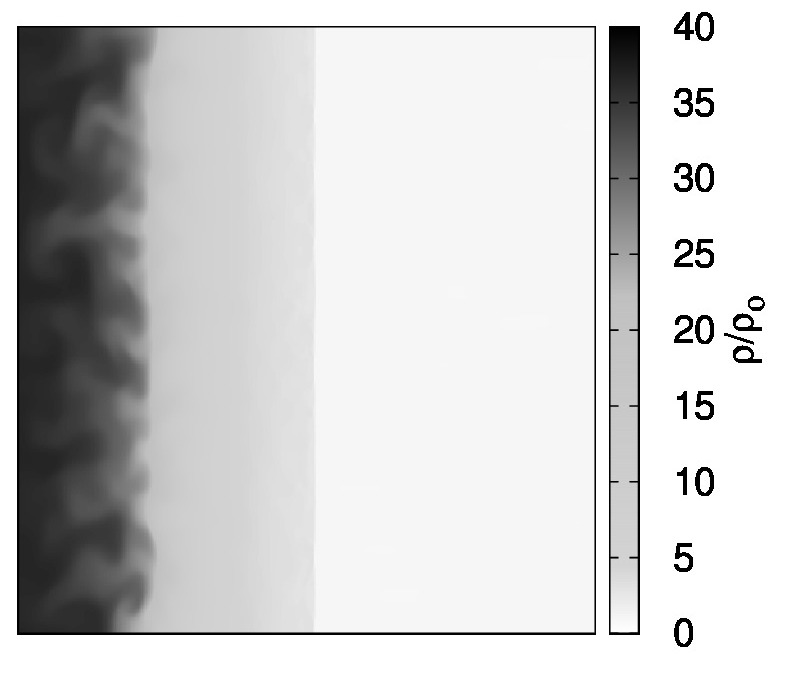}}\\
\caption{Comparison of the evolution of shock morphology from MD (top) and continuum (bottom) models for $u_p=20$, $u^*=10$ and $\varepsilon=0.95$.}
\label{fig:compareMDHD2D}
\end{figure}

Figure \ref{fig:compareMDHD2D} shows the evolution of shock morphology obtained via the MD and continuum models for $u_p=20$, $u^*=10$, and $\varepsilon=0.95$. Both models demonstrate an irregular shock structure with high density non-uniformities extending ahead of the piston. The continuum model shows a higher frequency instability, evident by the increased frequency of bumps. 

\begin{figure}[h]
\centering
\captionsetup[subfigure]{labelformat=empty, oneside,margin={1cm,.5cm}} 
\subfloat{\includegraphics[width=0.3\linewidth]{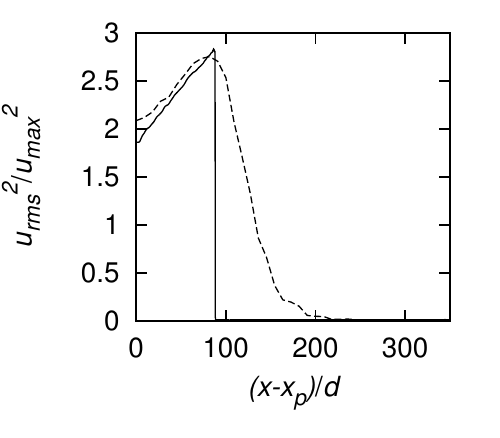}}\hspace*{-26.2pt}
\subfloat{\includegraphics[width=0.3\linewidth]{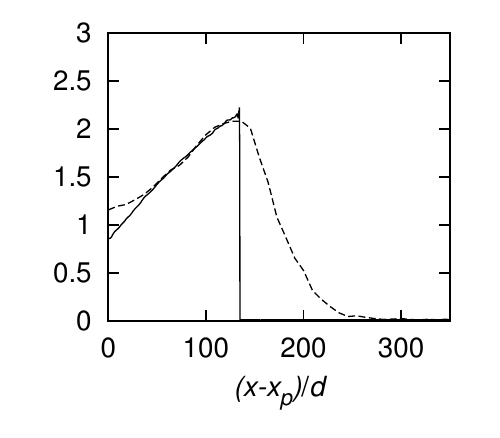}}\hspace*{-26.2pt}
\subfloat{\includegraphics[width=0.3\linewidth]{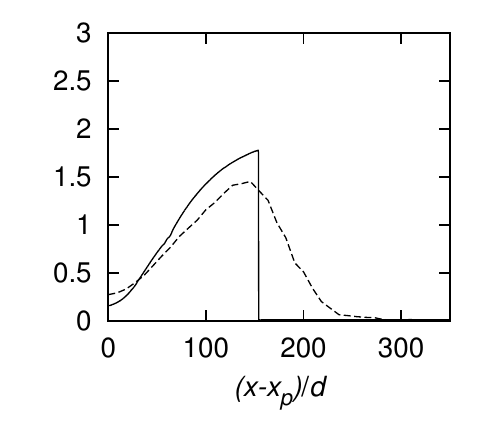}}\hspace*{-26.2pt}
\subfloat{\includegraphics[width=0.3\linewidth]{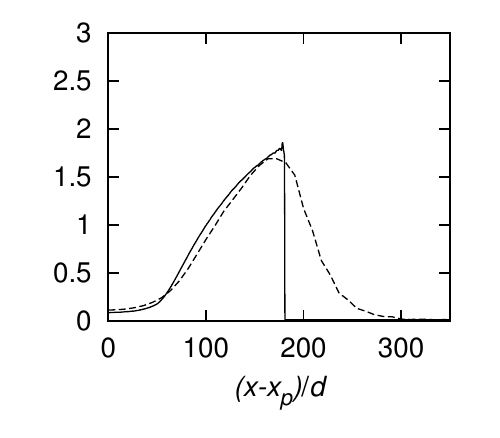}}\
\subfloat[$\overline{t}=11.15$]{\includegraphics[width=0.3\linewidth]{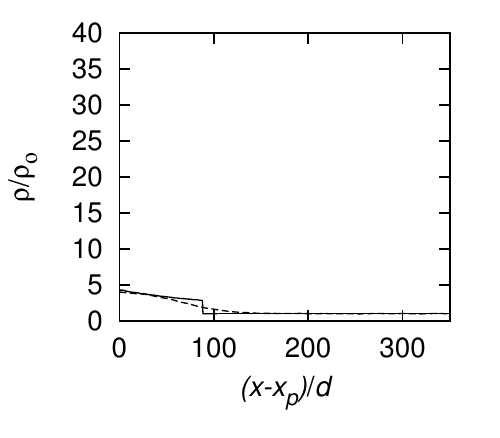}}\hspace*{-25.2pt}
\subfloat[$\overline{t}=22.3$]{\includegraphics[width=0.3\linewidth]{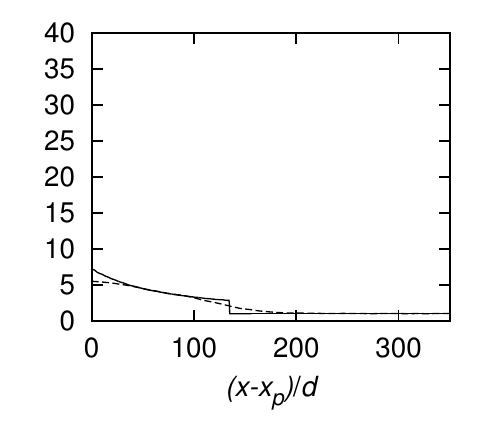}}\hspace*{-25.2pt}
\subfloat[$\overline{t}=55.75$]{\includegraphics[width=0.3\linewidth]{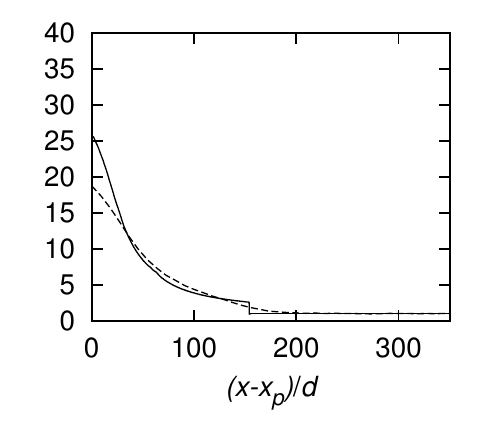}}\hspace*{-25.2pt}
\subfloat[$\overline{t}=111.5$]{\includegraphics[width=0.3\linewidth]{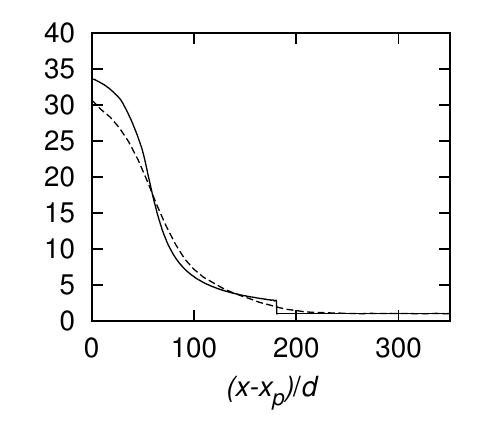}}
\caption{Evolution of one-dimensional temperature (top) and density (bottom) distributions, comparing MD (dashed) and continuum (solid) models for $u_p=20$, $u^*=10$ and $\varepsilon=0.95$.}
\label{fig:distributions}
\end{figure}

Figure \ref{fig:distributions} shows the evolution of the macroscopic 1D distribution of granular temperature and density, for the case seen in Figure \ref{fig:compareMDHD2D}. The MD results are obtained by ensemble and course grain averaging over 50 simulations with statistically different initial conditions. The granular temperature $u_{rms}^2$, is scaled by the activation energy $u_{max}^2$, which was previously found to form identical distributions for equal values of $u_p/u_{max}$ and $\varepsilon$ \cite{SirmasTBD}. 

The results for temperature and density distributions from the MD and continuum models show good agreement with regards to shock structure. Due to the lack of viscous effects in the continuum model, the shock front is represented by a sharp discontinuity on the temperature and density distributions. The locations of the peak temperature are similar for the two models, demonstrating that the shock waves are travelling at the same velocity.  

\section{Conclusions}
In this study, we have demonstrated that shock instability through granular gases can be seen at the discrete and continuum levels. Although we do not model viscous effects in the continuum model, instability is still present, albeit at a higher frequency than that seen in the MD simulations. These results may shed light on the role higher order effects have on the overall mechanism controlling shock instabilities in dissipative gases, which will be a subject of further investigation.
 
\bibliographystyle{iopart-num} 
\bibliography{references}

\end{document}